\documentstyle[aps,prl,epsfig]{revtex}

\begin{document}

\draft \small \twocolumn[\hsize\textwidth\columnwidth\hsize\csname
@twocolumnfalse\endcsname \title{Stable and unstable vortices in multicomponent
  Bose-Einstein condensates}

\author{Juan J. Garc\'{\i}a-Ripoll and V\'{\i}ctor M.
  P\'erez-Garc\'{\i}a,}

\address{ Departamento de Matem\'aticas, Escuela T\'ecnica Superior de
  Ingenieros Industriales\\ Universidad de Castilla-La Mancha, 13071 Ciudad
  Real, Spain }

\date{\today}

\maketitle

%%%%%%%%%%%%%%%%%%%%%%%%%%%%%%%%%%%%%%%%%%%%%%%%%%%%%%%%%%%%%%%%%%%

\begin{abstract}
  We study the stability and dynamics of vortices in two-species condensates as
  prepared in the recent JILA experiment ( M. R. Andrews {\em et al.}, Phys.
  Rev. Lett. 83 (1999) 2498).  We find that of the two available
  configurations, in which one specie has vorticity $m=1$ and the other one has
  $m=0$, only one is linearly stable, which agrees with the experimental
  results.  However, it is found that in the unstable case the vortex is not
  destroyed by the instability, but may be transfered from one specie to the
  other or display complex spatiotemporal dynamics.
\end{abstract}

\pacs{PACS number(s): 03.75. Fi, 67.57.Fg, 67.90.+z} ]

%03.75.Fi Phase coherent atomic ensemble (Bose condensation)
%67.57.Fg  Textures and vortices (superfluids)
%67.90.+z Other topics in quantum fluids and solids

\narrowtext

% -------------------------------------------------------------

Vortices appear in many different physical contexts ranging from classical
phenomena such as fluid mechanics \cite{Fluids} and nonlinear Optics
\cite{Kivshar} to purely quantum phenomena such as superconductivity and
superfluidity \cite{Superflu}.  In the last two years more than 100 papers
concerning vortices have been published in Physical Review Letters, which is a
naive  way to appreciate the importance of this subject in Physics.

A vortex is the simplest topological defect one can construct: in a closed path
around a vortex, the phase of the field undergoes a $2\pi$ winding and
stabilizes a zero value of the field placed in the vortex core.  The vortex is
stable because of topological constraints; removing the phase singularity
implies an effect on the boundaries of the system which cannot be achieved
using local perturbations.

Vortices are central to our understanding of superfluidity and quantized flow.
This is why after the experimental realization of Bose-Einstein condensates
(BEC) with ultracold atomic gases \cite{experim} the question of whether atomic
BEC's are superfluids has triggered the analysis of vortices.  The main goals
up to now have been to propose a robust mechanism to generate
\cite{generate,Ballagh} and detect vortices \cite{detect}. But another
important research area is the analysis of vortex stability
\cite{stability,Rokhsar,Vortices}, to which this works contributes.

Although most of the theoretical effort concerning vortices has been focused on
single component condensates, the first experimental production of vortices in
a BEC was attained using a two-species ${}^{87}$Rb condensate \cite{MMatt99}.
In this experiment the condensed cloud is made up of atoms in two different
hyperfine levels, denoted by $|1\rangle$ and $|2\rangle$. Since the scattering
lengths are different both states are not equivalent. As a consequence, while
each specie can host a vortex, in Ref. \cite{MMatt99} it was shown that a
single vortex is stable only when it is placed in the component with the
largest scattering length, $|1\rangle$. The other possibility, which has the
vortex in $|2\rangle$, leads to some kind of instability.

Our intention is to prove in this paper that the instability is purely
dynamical and can be explained with a mean field model which does not include
any type or dissipation. We will achieve this goal in three major steps. First,
we propose a model based on coupled Gross-Pitaevskii equations and solve the
stationary equations in two and three-dimensional setups. We obtain the lowest
energy stationary states that can be qualitatively identified with the ground
state and the two realizations of Ref. \cite{MMatt99}. Next we study the
stability of each state under small perturbations using linear perturbation
theory. Our main result is that only the experimentally stable configuration is
also linearly stable.  Finally, using numerical simulations, we study more
realistic conditions in which the condensate suffers moderate to strong
perturbations.  We show that there is a good agreement with experiment and also
that the dynamics is very rich and depends on the dimensionality of the system
and the intensity of the perturbations.

% ------------------------------------------------------------

{\em The model.-} In this work we will use the zero temperature approximation,
in which collisions between the condensed and non condensed atomic clouds are
neglected. In the two species case this leads to a pair of coupled
Gross-Pitaevskii equations (GPE)
\begin{mathletters}
\label{GPE}
\begin{eqnarray}
i \hbar \frac{\partial}{\partial t} \psi_1 &=& \left[ - \frac{\hbar^2\nabla^2}{2m} + 
 V_1 + \tilde u_{11} |\psi_1|^2 + \tilde u_{12} |\psi_2|^2 
 \right] \psi_1, \\
i \hbar \frac{\partial}{\partial t} \psi_2 &=& \left[ - \frac{\hbar^2\nabla^2}{2m} +
  V_2 + \tilde u_{21} |\psi_1|^2 + \tilde u_{22} |\psi_2|^2 
 \right] \psi_2.
\end{eqnarray}
\end{mathletters}
where $\tilde u_{ij} = \frac{4\pi\hbar^2}{m} a_{ij}$, and $a_{ij}$ are the
corresponding scattering length.  To simplify the formalism and in analogy with
experiments we assume that both potentials are spherically symmetric, i. e.
$V_1(\vec{r}) = V_2(\vec{r}) = \frac{1}{2}m\omega^2 (r^2+ z^2).$

Following the experiments, we will present results for equal number of particle
on each specie, $N_1=N_2=N$, which translates to
\begin{equation}
\int|\psi_1(\vec{r})|^2 d \vec{r} =\int|\psi_2(\vec{r})|^2 d\vec{r} \equiv 1.
\end{equation}
after a proper rescaling of $\psi_1$ and $\psi_2$ by $N$. To simplify the
analysis we change to a new set of units based on the trap characteristic
length, $a_0=\sqrt{\hbar/m\omega}$, and period, $\tau=1/\omega$. In this set of
units the nonlinear coefficients are $u_{ij} = 4\pi a_{ij} \sqrt{N_i N_j} /
a_0$. For the JILA experiment, in which $\omega=2 \pi \times 7.8\pm0.1 Hz$, we
have that $\tau\simeq 20.4 \ \hbox{ms}$ is the new unit of time.

The whole study, including the linear stability analysis and the numerical
simulations, was performed in two- and three-dimensional systems. We have
studied the system up to values of $u_{ij} \leq 5000$, which are of the order
of magnitude of the experiments.  Nevertheless, the linear stability analysis
and the simulations change little for the strongest interactions and we expect
our results to be still valid for a larger number of particles ($N \gg 10^6$).
Regarding the intensity of the nonlinearity, we have used the scattering
lengths of $\,^{87}$Rb which appear in Ref. \cite{Scattering-lengths}. These
values give us a precise line in the parameter space
\begin{equation}
\label{lengths}
U = g \left(\begin{array}{cc}1.00 & 0.97 \\ 0.97 & 0.94\end{array}\right).
\end{equation}
It is remarkable that because of the relation $u_{11} > u_{12} > u_{22}$ the
experiment is performed in a regime in which the first component chases the
second one, which rejects mixing with the chaser. This means that a ``desired''
configuration has the first component spread over the largest part of the
space. As we will see, this has important consequences for the dynamic of the
states.

% ------------------------------------------------------------

{\em Search of solutions.-} We are interested in stationary configurations in
which each component has a well defined value of the angular momentum. Such
states have the time and angular dependence factored out
\begin{equation}
\psi_i(r,z,\theta)=e^{-i\mu_it} e^{im_i\theta}\phi_i(r,z).
\end{equation}
These functions satisfy a nonlinear set of coupled PDE
\begin{equation}
\label{stationary}
\mu_t \psi_i = -\frac{1}{2}\nabla^2\psi_i+\frac{1}{2}\left(r^2+z^2\right) \psi_i+
\sum_{j}u_{ij}|\psi_j|^2\psi_i,
\end{equation}
with $i=1,2$. Our focus will be on three particular configurations, which are
the lowest energy states with vorticities $(m_1,m_2)=(0,0),(1,0),(0,1)$. They
correspond to the ground state of the double condensate, and to the single
vortex states for the $|1\rangle$ and $|2\rangle$ species, respectively.

\begin{figure}[p]
\begin{center}
  \epsfig{file=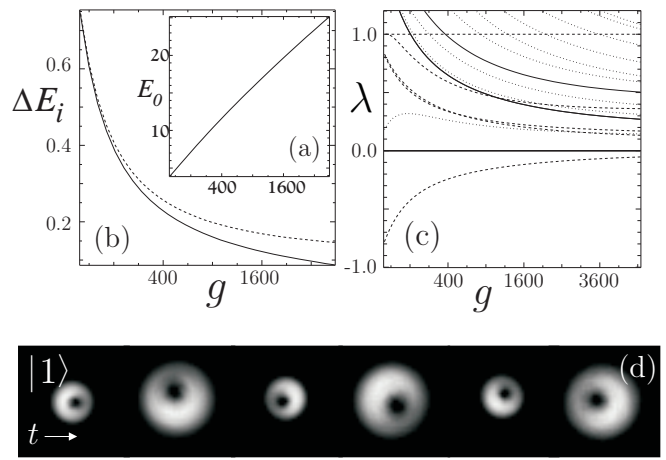,width=1\linewidth,clip=1}
\end{center}
\caption{
  (a) Energy of the ground state, $(0,0)$, (b) energy of the $(1,0)$ (solid
  line) and $(0,1)$ (dashed line) configurations relative to the ground state
  and (c) frequencies of the linear modes around the $(1,0)$ stable state, with
  respect to the nonlinearity, given by $g$ [Eq. (\ref{lengths})]. (d) Upper
  view of component $|1\rangle$ with a vortex after a strong perturbation of
  the cloud (see above) for $g=1500$. Pictures are equispaced by 10 time units
  and each one is $18\times 18$ spatial units large.}
\label{fig-stable}
\end{figure}

To find the radial and longitudinal dependences of the wave functions,
$\phi_i(r,z)$, we expand them approximately on a finite subset of the harmonic
oscillator basis, $\phi_i(r,z) \simeq \sum_{n=0}^{n=N} c_n P_n^{(m_i)}(r,z)
e^{-r^2/2} e^{-z^2/2}$ and then search the ground states for each of the
$(m_1,m_2)$ pairs of vorticities. The details of the method applied to single
vortex systems can be found in Ref. \cite{Vortices}. As a result one obtains
the desired eigenfunctions, chemical potential and energies [Fig.
\ref{fig-stable}(a)] of each $(m1,m2)$ ground state.

{\em Linear stability analysis.-} Let us study the behavior of the stationary
states under infinitesimal perturbations (e.g. in the initial data, small
amounts of noise, etc.).  To do so, we define the excitations as
\begin{mathletters}
\label{mate_a}
\begin{eqnarray}
\psi_i(r,z,\theta) & = &e^{-i\mu_it + im_i\theta}\left(\phi_i(r,z)+e^{-i\lambda
t+in\theta}\alpha_i(r,z)\right),  \\
\bar{\psi}_i(r,z,\theta) & = &e^{i\mu_it - im_i\theta}\left(\bar{\phi}_i(r,z)+e^{-i\lambda
t-in\theta}\beta_i(r,z)\right).
\end{eqnarray}
\end{mathletters}
Then we introduce Eqs. (\ref{mate_a}) into Eq. (\ref{GPE}) and keep the ${\cal
  O}(\alpha)$ and ${\cal O}(\beta)$ terms. In the end we reach an equation for
the perturbations $\vec{W} = [\alpha, \alpha^*, \beta, \beta^*]$ which is of
the type $iJ\partial_t\vec{W} = H\vec{W}$. Here $H$ is an hermitian operator
and $iJ$ is an anti-hermitian operator. We will search a Jordan basis such that
$\lambda\vec{W}=-iJH\vec{W}.$ The lack of such a Jordan form leads to
polynomial instabilities, while the presence of non-real eigenvalues is a
signal of the exponentially growing instabilities. All other modes give us the
frequencies of the linear response of the system to small perturbations.

This analysis is formally equivalent to the Bogoliubov stability analysis of
the states under consideration. However, as it was shown in \cite{Vortices},
our perturbation is of order ${\cal O}(\alpha^2,\beta^2)$ in the energy and thus the
exponentially unstable modes lay along directions of constant energy and thus a
Bogoliubov expansion of the Hamiltonian cannot account for the instability,
which can only be reached by working directly on the GPE. It is thus important
to confirm our linear stability results with numerical simulations of the whole
system.

{\em Linear stability results}.- When we perform the diagonalization on
$(m_1,m_2)=(0,0)$ we obtain that all of the eigenvalues are positive numbers,
which implies that $(0,0)$ is at least a local minimum of our energy functional
--furthermore, it is the ground state of the system and thus globally stable.

Next we studied the $(1,0)$ family [Fig. \ref{fig-stable}(c)] and find that
there is a negative eigenvalue among an infinite number of positive ones, which
means that there is a single path in the configuration space along which the
energy decreases. That direction belongs to a $m=0$ perturbation which takes
the vortex out of the condensate \cite{Rokhsar}.  Nevertheless, as in this case
there exist no complex eigenvalues, we conclude that the lifetime of the
configuration is only limited by the presence and amount of the losses. This is
confirmed when take a $(1,0)$ configuration, perturb it and study its real time
evolution [Fig. \ref{fig-stable}(d)], and it is indeed consistent with the
experiments of Ref. \cite{MMatt99}.

We must remark that the existence of a negative eigenvalue in the spectrum
around the $(1,0)$ vortex contradicts the belief that the second component
could act as a pinning potential. Indeed, based on further study we can state
that a $(1,0)$ vortex remains energetically unstable even for different
proportions of each specie, from $N_1 \simeq 0$ to $N_2 \simeq 0$
\cite{two-modes}.

\begin{figure}[p]
\begin{center}
  \epsfig{file=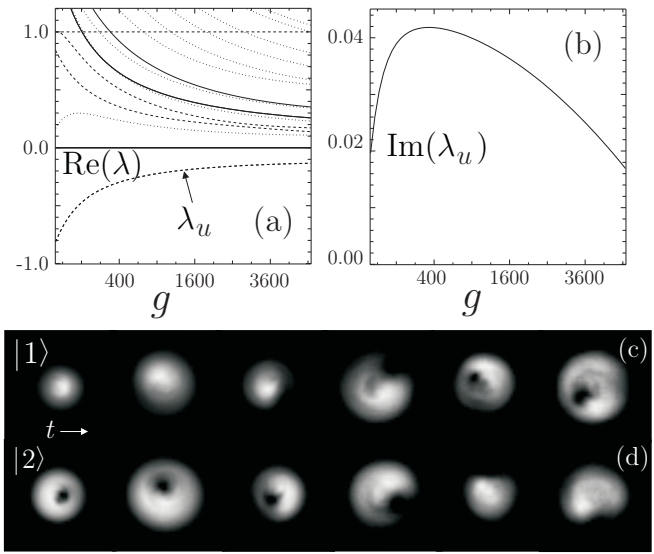,width=1\linewidth,clip=1}
\end{center}
\caption{(a) Linear response frequencies around the
  $(0,1)$ unstable state and (b) imaginary part of the unstable mode arising
  from the linear stability analysis, both with respect to the nonlinearity $g$
  [Eq. (\ref{lengths})] (c-d) Upper view of an unstable vortex after a finite
  perturbation for $g=1500$. Shown are species $|1\rangle$ (c) and $|2\rangle$
  (d). Pictures are equispaced by 10 time units and each one is $18\times 18$
  units large.}
\label{fig-unstable}
\end{figure}

Finally we focus on the $(0,1)$ family [Fig. \ref{fig-unstable}(a)]. Here we
find an infinite number of modes with positive energy, plus a pair of them with
negative frequency and complex eigenvalues, $\lambda_u$.  Qualitatively, the
shape and frequencies of the unstable modes are similar to those of the energy
decreasing modes of the $(1,0)$ --that is, they are perturbations which push
the vortex out of both clouds. The difference is that due to the imaginary part
of those eigenvalues, which is Im$(\lambda_u)={\cal O}(0.04)$ [Fig.
\ref{fig-unstable}(b)], vortices with unit charge in $|2\rangle$ are unstable
under a generic perturbation of the initial data.  This is consistent with the
JILA experiments, where the a vortex hosted inside a $|2\rangle$ specie was
found to be unstable.

Although, as we mentioned above, the previous result does not depend
drastically on $N_1$ being equal to $N_2$, we have also found that for $N_2 \gg
N_1$ the vortex in $|2\rangle$ becomes practically stable \cite{two-modes}
--that is, the lifetime is too long to be observed in numerical simulations.
This is consistent with the limit of a single-specie condensate where the
unit-charge vortex is found to be stable \cite{Vortices} for any scattering
length.

{\em Does the vortex break?.-} The linear stability analysis cannot be used to
draw conclusions about the behavior of the vortex far from the limit of
infinitesimal perturbations. To get further insight on the dynamics we have
performed a set of numerical simulations in which we reproduce different
realistic perturbations of each state. These perturbations vary from small
displacements of the vortex core which should agree in detail with the linear
stability results, to strong perturbations of the initial data. In particular,
we have systematically used the procedure of creating a stationary state and
then reducing the size of the trap, thus transfering the system to a non
stationary state \cite{why-oscillate}.  The numerical simulations have been
done using a symmetrized split-step Fourier pseudospectral method on grids of
sizes ranging from $32\times 32 \times 32$ to $64 \times 64 \times 64$ for the
three-dimensional setups, and $64 \times 64$ to $256 \times 256$ for the
two-dimensional model. All results were tested on different grid sizes and
changing time steps to ensure their validity.

From our numerical simulations we extract several conclusions. First, the
linearly stable state, $(1,0)$, is also a robust one and survives to an ample
range of perturbations, suffering at most a precession of the vortex core plus
changes of the shapes of both components [Fig. \ref{fig-stable}(d)].  This
behavior is equally reproduced in two- and three-dimensional simulations.

Next we have the unstable states, $(0,1)$ subject to weak perturbations in
either two- or three-dimensional models. In that case the unstable
configuration develops a simple recurrent dynamics, which is well represented
by Figs. \ref{fig-unstable}(d-e). There we see a first stage where the first
component [Fig. \ref{fig-unstable}(d)] and the vortex [Fig.
\ref{fig-unstable}(e)] oscillate synchronously (the hole in $|2\rangle$ pins
the peak of $|1\rangle$).  These oscillations grow in amplitude until the
linked system spirals out and forms a ying-yang which rotates clockwise, one
specie chasing the other.  Finally the first component develops a tail and
later a hole which traps the second component.  That hole is a vortex, which
somehow has been transfered from $|2\rangle$ to $|1\rangle$.  Though it is not
completely periodic, this mechanism does exhibit some recurrence and the vortex
returns to $|2\rangle$.

\begin{figure}[p]
\begin{center}
  \epsfig{file=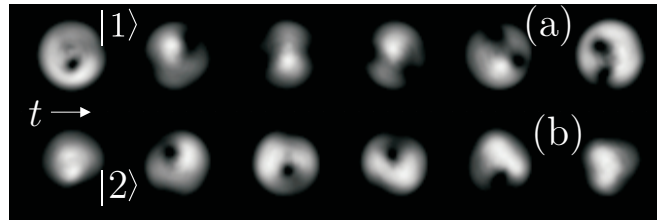,width=1\linewidth}
\end{center}
\caption{
  Evolution of an unstable $(0,1)$ configuration subject to a finite
  perturbation after a 50 units run for $g=1500$. Spatiotemporal chaos with
  several vortices develops for long enough times. Shown are upper views
  equispaced by 10 time units, each one $18\times 18$ spatial units large.}
\label{fig-turb}
\end{figure}

The preceding behavior persists even for strong perturbations in a
two-dimensional condensate. However, when one considers a three dimensional
condensate and applies large perturbations on the initial data such as the one
described above, it is possible to find a richer dynamics which include the
establishment of spatiotemporal chaos. As an example, Fig. \ref{fig-turb} shows
a regime in which more than one vortex are introduced into the first component
for long times. Intuitively, this turbulent behavior has two causes.  First,
due to energetic considerations there is a bigger overlap of both species than
in the two dimensional model, as is apparent from the pictures. And second and
most important, in the a three dimensional environment the first component is
more reluctant to be dragged by the weak vortex-line from the second component.
Thus, as the second component spirals around, it is able to shake the first
fluid and produce pairs of vortices, much like what happens in laser-stirred
condensates \cite{Ballagh}.

\begin{figure}[p]
\begin{center}
  \epsfig{file=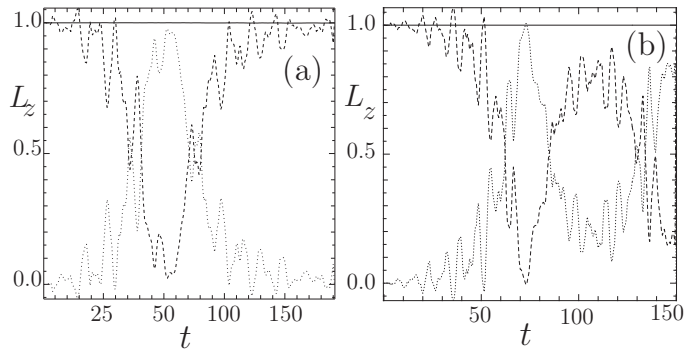,width=9cm}
\end{center}
\caption{Evolution of total angular momenta $L_z$ (solid line, well conserved by our
  numerical scheme), and partial angular momenta $L_z^{(1)}$ (dotted line) and
  $L_z^{(2)}$ (dashed line), with respect to time in adimensional units.  Plot
  (a) belongs to a situation with a simple, recurrent transfer of the vortex,
  while (b) corresponds to a chaotic spatiotemporal dynamics [Fig.
  \ref{fig-turb}].}
\label{fig-lz}
\end{figure}

Either phenomena, the transfer of the vortex and the chaotic evolution do not
not contradict any topological rule or conservation law. In fact the angular
momentum of each component is no longer a conserved quantity, and the
topological charge of each specie needs not survive through the evolution.
Instead, what is conserved is the total angular momentum
\begin{equation}
L_z =  i\hbar\int {\bar{\psi}}_1\partial_\theta \psi_1 +
       i\hbar\int {\bar{\psi}}_2\partial_\theta \psi_2=L_z^{(1)} + L_z^{(2)}.
\end{equation}
In Fig. \ref{fig-lz} we plot the evolution of the total angular momentum and
that of each component $L_z^{(j)}$ for a recurrent situation and for the case
of a spatiotemporally chaotic state [Fig. \ref{fig-turb}].  It can be seen that
there is a complex interchange of angular momentum between both components. The
intermediate states are topologically nontrivial ones since the phase
singularity is being transfered from one component to the other.  A more
detailed analysis of this process will be presented elsewhere \cite{two-modes}.

{\em Conclusions and discussion.-} We have analyzed the stability of vortices
in multi-component atomic Bose-Einstein condensates, using both two-dimensional
and three-dimensional sets of coupled Gross-Pitaevskii equations.  We prove
that a vortex in a $|1\rangle$ state is a dynamically stable object even though
it is not a global minimum of the energy.  In contrast, we demonstrate that a
vortex in the $|2\rangle$ specie is dynamically and energetically unstable and
tends to spiral out of the condensate. Besides, since our model does not
involve dissipative effects or asymmetries, one cannot get rid of the vortex
angular momentum. This leads to a complex dynamics in which the vortex is
transfered to the $|1\rangle$ state and eventually goes back to $|2\rangle$.
This and other predictions about the dynamical behavior can be checked in
current experiments.

We believe that the simple model presented in this paper gives a reasonable
explanation of the experiments of \cite{MMatt99} based only on the nonlinear
interactions present in mean field theories.  In fact if the instability found
in the experiments were due to dissipation through a mechanism similar to that
proposed in Ref.  \cite{Schlyapnikov} it would affect both the $(1,0)$ and
$(0,1)$ type of vortices in a similar way, which is not the case.  In support
to our theory we can see from Fig. 3 of Ref.  \cite{MMatt99} that the vortex
does not completely escape but some kind of defect is formed in the periphery
of the $|2\rangle$ component, a behavior similar to that found in our real time
simulations [Fig 2(c)].

Regarding the time scales, our {\em linear} stability analysis gives lifetimes
of about 500 milliseconds, which is two or three times larger than what is seen
in the experiments. Nevertheless this is not significant since further study
has revealed that the time after which the instability affects the system 
depends dramatically on the type and the intensity of the perturbation, which 
the linear stability analysis requires to be small. As such it is conceivable
that the experimental realization of the $(0,1)$ vortex must break sooner: the 
whole experimental preparation takes the system to a state which differs 
{\em finitely} from the stationary configuration and are thus more likely to 
excite the unstable mode, even through the preparation process.

This work has been partially supported by the DGICYT under grant PB96-0534.

% --------------------------------------------------------------


\begin{references}
\bibitem{why-oscillate}{This is why the, in Figs. \ref{fig-stable}(e) and
    \ref{fig-unstable}(c), oscillations of the whole cloud are seen in addition
    to the relevant stable or unstable dynamics.}
\end{references}
\end{document}